\documentclass[a4paper]{article}
\usepackage{amsfonts}
\usepackage{amssymb}
\usepackage[margin=2cm]{geometry}
\usepackage[dvips]{graphicx}

\usepackage[T1]{fontenc}
\usepackage[latin2]{inputenc}

\usepackage{amsmath}
\usepackage{bbm}

\begin{document}


\title{Physical limits on self-replication processes}

\author{Robert Alicki\\
 Institute of Theoretical Physics and Astrophysics, University
of Gda\'nsk, \\ Wita Stwosza 57, PL 80-952 Gda\'nsk, Poland}

\date{\today}

\maketitle

\begin{abstract}
Using few very general axioms which should be satisfied by any reasonable theory consistent with general physical principles and some more recent
results concerning "broadcasting" of quantum states we show that: a) only classical information can self-replicate perfectly, b) "parent" and "offspring" must be strongly correlated, c) "separation of species" is possible only in a non-homogeneous environment. To illustrate the existence of theoretical schemes which possess both classical and quantum features, we present a model based on the classical probability but with overlapping pure states and "entangled states" for composite systems. 

\end{abstract}

In his essay from 1967  Wigner \cite{Wig}  argued that the phenomenon of self-replication of biological molecules  and organisms contradicted the principles of quantum mechanics.  In 1971  Eigen \cite{Eig}  responded  with an argument that the Wigner choice of  a typical (random) unitary map as a quantum dynamics did not take into account the instructive functions of informational macromolecules.  In 1982 Wooters,  Zurek  \cite{Woo} and   Dieks 
\cite{D}, apparently not awared about the previous debate,   proved  the Wigner "no-cloning theorem"  for an arbitrary quantum unitary dynamics what made Eigen's criticism invalid. Motivated by the importance of the {\sl no-cloning theorem} in modern quantum information theory we revisit the problem stated by Wigner and discuss the limitations on self-replication phenomena which follow from the very general physical principles, in particular the second law of thermodynamics.

We begin with the  Wigner formulation of  the self-replication model denoting by $\phi$ the state of  an "organism" and by $\omega $   the fixed initial state of "food" (environment).  He considered a self-replication process  as the following dynamical transformation from the initial state to the final one
\begin{equation}
\phi\otimes\omega \mapsto \phi\otimes\phi\otimes\sigma = T(\phi\otimes\omega)
\label{srW}
\end{equation}                                     
where $\sigma$ is the final state of  environment which may depend on the state $\phi$,  and $T$ denotes the dynamics defined on the total closed system. The symbol $\phi\otimes\psi$ denotes an abstract product state for a system composed of two statistically independent subsystems.

In contrast to the previous works  dealing with quantum mechanical models \cite{Wig,Woo,D,H3,Bar}, we assume that the theory of self-replication processes  satisfies only few very general axioms:
   
A1)  For any two states  $\phi$  and  $\psi$ of  a system there exists an "overlap"  $(\phi | \psi )$ which measures the {\sl indistinguishability} of two states  and satisfies the conditions
\begin{equation}
0  \leq    (\phi| \psi)  \leq 1\  ,  (\phi| \psi ) = 1\   {\rm if\ and\ only\ if}\     \phi = \psi .
\label{a1}
\end{equation}
A2) For all product states $\phi\otimes\psi$, $\phi'\otimes\psi'$  the following factorization holds
\begin{equation}
( \phi\otimes\psi | \phi'\otimes\psi')  = (\phi | \phi' ) (\psi |\psi') . 
\label{a2}
\end{equation}                                      
A3) Any dynamics of  a {\sl closed system} given by a map $\phi\mapsto T(\phi)$ does not reduce the overlap of two arbitrary states 
\begin{equation}
(T(\phi) | T(\psi)) \geq  (\phi |\psi )   .
\label{a3}
\end{equation}                                                        
A4) Joint states $\Phi$ ,$\Psi$ of a composed system  are better distinguishable than the corresponding reduced states of subsystems , $\phi_1 , \psi_1$  and  $\phi_2 , \psi_2$, i.e. 
\begin{equation}
(\phi_1 |\psi_1)\   {\rm and}\   (\phi_2 |\psi_2) \geq   (\Phi |\Psi)   . 
\label{a4}
\end{equation}                                                                                 
Both,  classical and quantum statistical mechanics  fulfil the Axioms A1)-A4) with the choice 

$(P | P') =\int\sqrt{p(x)p'(x) } dx $  for classical probability  distributions and   
$(\rho | \rho')  = {\rm Tr} (\sqrt{\sqrt{\rho}\rho'\sqrt{\rho}} )$  for  quantum mechanical density matrices, respectively \cite{Bar}. Moreover, for classical and quantum Hamiltonian dynamics we have always the equality in (\ref{a3}). A general inequality (\ref{a3}) can be treated as a form of the Second Law of Thermodynamics for closed systems - {\sl information about any closed system does not increase during the evolution}. Indeed, the decrease of an overlap between
two states means better distinguishability of them and therefore an information gain.

Consider now a family of  "species"  described by the states $\phi ,\phi',\phi'',...$ Their self -replication is governed by the cloning process  (\ref{srW}) with the dynamics T and the fixed initial state of  an environment $\omega$. We call two states $\phi$ and $\psi$  {\sl disjoint}  if  $(\phi | \psi) = 0$.  The general "no-cloning theorem" says  that:

{\sl The cloning process is only possible for pairwise disjoint states  .}

{\sl Proof}.  This theorem follows from the axioms A1)-A3) only. Namely, 
\begin{equation}
(\phi |\phi') =  (\phi\otimes\omega | \phi'\otimes\omega ) \leq  ( T(\phi\otimes\omega)| T(\phi'\otimes\omega) ) 
 =  (\phi\otimes\phi\otimes\sigma |\phi'\otimes\phi'\otimes\sigma' )    = (\phi| \phi')^2  (\sigma |\sigma')
 \leq (\phi| \phi')^2         
\label{ncp}
\end{equation}                                                                                     
and hence   either  $\phi = \phi'$  or   $(\phi |\phi') = 0$.

Different species are described by different  states of  a certain complex system identified, for example, with the probability distributions (or their quantum counterparts - density matrices) over the family of relevant parameters. Such states represent corresponding ensembles of individual biological molecules , organisms, etc.  During the slow evolution process those "species-states" may become more or less distinguishable, but generally, there are no reasons to assume that at the given moment they all are disjoint.  On the other hand a single step of self-replication should be treated as an almost perfect one.  As a consequence, the general no-cloning theorem implies that the Wigner formula (\ref{srW}) cannot provide a proper generic model of self-replication. In particular, the product structure $\phi\otimes\phi$ of the joint state of parent and offspring is a too restrictive assumption.

Indeed, a more general scheme studied  by Barnum et.al.\cite{Bar}  in the context of quantum information and called "broadcasting" represents better the concept of self-replication.  Consider the following generalization of  (\ref{srW})
\begin{equation}
\phi\otimes\omega\mapsto \Phi\otimes\sigma = T(\phi\otimes\omega)
\label{bro}
\end{equation}       
where now $\Phi$ is a joint state of  parent and offspring.  Here again, $\omega$ is a fixed initial state of an environment and the final state $\sigma$ may depend on $\phi$. From (\ref{bro}) and the axioms A1)-A4)  the following inequalities follow:
\begin{equation}
(\phi_1|\phi'_1)\ {\rm and}\ (\phi_2|\phi'_2) \geq (\Phi |\Phi') \geq (\phi|\phi')\ .
\label{bro1}
\end{equation}  

Here  $\phi_1 , \phi'_1$,  and  $\phi_2 , \phi'_2$  are reduced states describing parent and offspring after self-replication, respectively.

Assuming the natural condition, that the state of  parent does not change in the self-replication process, i.e. $\phi_1 = \phi$ , $\phi'_1 = \phi'$ , we obtain using again A1)-A4) and (\ref{bro1})
\begin{equation}
(\phi_2| \phi'_2)\geq (\phi_1 |\phi'_1)=(\phi |\phi')\ {\rm and} \   (\Phi | \Phi')  =  (\phi | \phi')\ .
\label{par1}
\end{equation} 
The first inequality shows that the overlap for  offspring  states is never smaller than that for  parent ones, what means that the process (\ref{bro}) typically leads to a "convergence of species".  A "separation of species" can be consistent with the self-replication process (\ref{bro}) only if the initial states of  environment $\{\omega\}$ are different for the different initial states of parent, {\sl i.e. when the environment is non-homogeneous}. 

A perfect  broadcasting takes place if also the states of  offspring are perfectly reproduced  states of  parent, i.e. $\phi_2 = \phi$ , $\phi'_2 = \phi'$.  Barnum et.al. \cite{Bar}  proved that in the quantum theory a perfect broadcasting is possible only for a set of quantum states given by pairwise commuting density matrices. Such states are essentially "classical" and can be represented by a family of probability distributions over the joint space of parameters. As a consequence, {\sl only  classical information can self-replicate perfectly}.

 In the classical theory a perfect broadcasting is possible but the condition $(\Phi | \Phi')  =  (\phi | \phi')$ (\ref{par1}) implies {\sl very strong correlations between parent and offspring}.  As an example one can take a perfect deterministic broadcasting of any  probability distribution $ p(i) \mapsto P(i,j) = p(i)\delta_{ij}$. Then, indeed  $(P| P') = (p | p')$. 

The usual interpretation of the no-cloning theorem underlines the difference between quantum and classical case. Namely, for quantum systems different pure states may overlap while in the classical theory pure states are always disjoint and therefore the no-cloning theorem seems to be interesting for quantum systems only. However, as argued above, 
the reduced  probabilistic ("fuzzy") description of complex systems typically involves overlapping probability distributions corresponding to different categories of objects ("species"). Such a fuzzy description can be formalised by assuming certain intrinsic indeterminacy relations which introduce "quantum features" into the classical theory.

As an illustration, consider a toy model in which all systems are described by
discrete probability distributions. We assume, however, that the detailed knowledge about any system is
forbidden. Namely, we impose the following condition for the set
${\cal S}_A$ of all states for any system $A$
\begin{equation}
p^A\equiv\{p_1 , p_2,...,p_{n_A}\} \in {\cal S}_A \ {\rm if}\
S(p)\equiv -\sum p_j\log p_j \geq \epsilon >0 \label{cond}
\end{equation}
where $\epsilon$ is a certain universal constant equal
for all systems and characterizing the intrinsic indeterminacy of all states. Obviously, ${\cal S}_A$ is a convex set. Extreme
points of this set, pure states, are those with the entropy equal
to $\epsilon$. Generally, such pure states {\sl need not to be
disjoint}.

For a system composed of two subsystems $A$ and $B$ we can define
a set of {\sl separable states} as all convex combinations of the
product states
\begin{equation}
{\cal S}^{sep}_{AB}= \bigl\{ p^{AB}=  \sum \lambda_j
p^{(A,j)}\otimes p^{(B,j)};p^{(A,j)}\in{\cal S}_A\ ,\  p^{(B,j)}\in{\cal S}_B\bigr\}\ . \label{sep}
\end{equation}
Obviously, {\sl non-separable states} form a set ${\cal S}_{AB}\setminus{\cal S}^{sep}_{AB}$.

We can prove that:

{\sl Non-separable states form a proper generic
subset of all states.}

{\sl Proof.} It is enough to estimate the entropy of a separable
state
\begin{equation}
S( \sum \lambda_j p^{(A,j)}\otimes p^{(B,j)}) \geq \sum \lambda_j
S( p^{(A,j)}\otimes p^{(B,j)}) = \sum \lambda_j
[S( p^{(A,j)}) + S( p^{(B,j)})]\geq 2\epsilon\ . \label{sep1}
\end{equation}
Then the states with the entropy in the interval $[\epsilon, 2\epsilon)$ are non-separable ("entangled").
In contrast to quantum
mechanics, separable pure states do not exist within this model.

In conclussion one should notice that the Wigner scheme \cite{Wig} of the cloning process (\ref{srW}) is too narrow even under the assumtpion that only classical information should self-replicate. The results on broadcasting \cite{Bar} show that indeed only classical information can self-replicate perfectly and moreover the general axioms containing a form of the second law of thermodynamics imply quite severe restrictions on these processes. They involve strong correlations between parent and offspring and the general tendency to "convergence of spieces" which can be reversed in a non-homogeneous environment only. A toy model of complex systems with intrinsic indeterminacy shows that classical models can display some quantum features including "entanglement".

The author is grateful to Micha\l\  and Ryszard Horodecki, and Marco Piani for discussions.
The work is supported by
by the Polish Ministry of Science and Information
Technology - grant PBZ-MIN-008/P03/2003

\end{document}